# Effect of annealing on the magnetic, magnetocaloric and magnetoresistance properties of Ni-Co-Mn-Sb melt spun ribbons

Roshnee Sahoo[1], D. M. Raj Kumar[2], D. Arvindha Babu[2], K. G. Suresh[1*], A. K. Nigam[3] and M. Manivel Raja[2]

[1] Department of Physics, Indian Institute of Technology Bombay, Mumbai- 400076, India

[2] Defence Metallurgical Research Laboratory, Hyderabad – 500058, India

[3] Tata Institute of Fundamental Research, Mumbai- 400005, India

The structural, magnetic, magnetocaloric and magnetotransport properties of $Ni_{46}Co_4Mn_{38}Sb_{12}$ melt spun ribbons have been systematically investigated. The partially ordered B2 phase of the as-spun ribbon transforms to fully ordered $L2_1$ phase upon annealing, which signifies a considerable change of the atomic ordering in the system. The presence of atomic disorder in the as-spun ribbon gives rise to a higher martensitic transition temperature and a lower magnetization as compared to the bulk sample. However, annealing the ribbons helps in regaining the bulk properties to a large extent. Significant changes in magnetocaloric effect, exchange bias and magnetoresistance have been observed between the as-spun and the annealed ribbons, indicating the role of atomic ordering on the functional as well as fundamental properties in the Heusler system. Importantly, the study shows that one can reduce the hysteresis loss by preparing melt spun alloys and subjecting them to appropriate annealing conditions, which enable them to become practical magnetic refrigerants.

*Corresponding author (email: suresh@phy.iitb.ac.in, FAX: +91-22-25723480)

# 1. Introduction

The Ni-Mn based ferromagnetic Heusler alloys crystallize in $L2_1$ cubic Heusler structure at high temperatures, which transforms to tetragonal or orthorhombic structure through first order martensitic transition upon cooling. Associated with the martensitic transition, several important functional properties such as shape memory effect [1], inverse magnetocaloric effect [2, 3] and giant magnetoresistance [4, 5] have been reported in many full Heusler alloys. Hence, both from fundamental and functional aspects, these materials are very promising. The magnetism in these systems is mainly attributed to the localized Mn magnetic moment. The ferromagnetic (FM) ordering of Mn moment is usually interpreted via the indirect exchange interaction mediated by the conduction electrons.

Recently rapid quenching or melt spinning technique has been attempted in a few Heusler systems such as Ni-Fe-Ga, Ni-Mn-Ga, Ni-Mn-In and Ni-Mn-Al alloys [6-9]. The melt spinning technique is advantageous in the synthesis of strongly textured polycrystalline ribbons and certain ribbon shapes are useful for certain applications. Additional advantage of melt spun ribbons is that the thin ribbons show low eddy current losses and can be utilized at high frequencies. Most of the samples prepared by melt spinning do not solidify directly to $L2_1$ or highly stable/equilibrium phase; instead solidifies to B2 or partially disordered phase. As is well known, full Heusler alloys are in the form of $X_2YZ$, where X and Y are 3d transition elements. In rapidly solidified alloys in B2 phase, Y and Z atoms are evenly distributed. By subsequent annealing or heat treatment the alloy exhibits long range $L2_1$ ordering, where Y and Z occupy their corresponding sites. The atomic rearrangements of these atoms upon annealing modify the electronic structure and hence all related properties. There are various studies in Ni-Mn based systems where a correlation between atomic order and magnetic order has been extensively studied. It has been reported in NiMnGa system that after heat treatment, the equilibrium phase can be achieved, which is due to the diffusion of Mn atoms from Ga sublattice to their own sublattice.[10] In NiMnGa and NiFeGa systems, it has been established that a change in the degree of atomic order significantly affects the structural and magnetic properties [10, 11].

In $Ni_{50-x}Co_xMn_{38}Sb_{12}$ bulk alloys, various functional properties like magnetocaloric effect (MCE), exchange bias (EB) and magnetoresistance (MR) have been extensively studied, [12-14] where Co addition for Ni affects the martensitic characteristics as well as the magnetic properties. The martensitic transition temperature is critically dependent on the valence electrons per atom (e/a ratio). Thus, a change in the composition or stoichiometry is expected to alter the martensitic/magnetic transition temperatures of the alloys. It is well known that there are other physical parameters such as external field, hydrostatic pressure and stress that influence the martensitic transition. However, there is very little study on the effect of the changes in the atomic ordering on the magnetic and the related properties of this system. Among the members of (bulk) NiCoMnSb series, $Ni_{46}Co_4Mn_{38}Sb_{12}$ has been found to be the most promising. Therefore, we have carried out melt spinning of this alloy and studied the effect of annealing on the martensitic, magnetic, magnetocaloric and magneto-transport properties.

## 2. Experimental procedures

Polycrystalline sample of $Ni_{46}Co_4Mn_{38}Sb_{12}$ was prepared by arc melting in high purity argon atmosphere. The constituent elements were at least of 99.99% purity. The ingots were remelted several times and the weight loss after the final melting was negligible. About 5 g of the as-melted alloy was taken in a quartz tube with a 1 mm diameter nozzle and was induction melted in flowing argon. The molten alloy was rapidly quenched in a high vacuum melt spinner, which yielded the as-spun ribbon samples of typically of 3-5 mm width and 20-30 μm thickness. Annealing was performed for some ribbons at 800 °C for 1 hr in argon atmosphere followed by furnace cooling. The structural characterization was done by powder x-ray diffraction (XRD) using Cu-Kα radiation. Microstructure of the melt spun ribbons was examined using high resolution field emission gun transmission electron microscope (FEG-TEM). The FEG-TEM was operated with 200 keV. For using TEM, thinning of the ribbon was done using PIPS (precision ion polishing system). The sample was milled gently using 5 kV in varying angles with ion gun. The sample composition analysis was performed in energy dispersive x-ray analysis (EDS) attached to SEM instrument. The magnetization measurements were carried out using a vibrating sample magnetometer attached to a Physical Property Measurement System (Quantum Design, PPMS-6500) and the electrical resistivity measurements were performed by four probe method using PPMS.

## 3. Results and discussions

Figure 1(a) and (b) show the room temperature XRD patterns of the as-spun and the annealed $Ni_{46}Co_4Mn_{38}Sb_{12}$ ribbons, respectively. The as-spun ribbon exhibits austenite phase with B2 ordering at room temperature. It can be seen that the odd (h k l) peaks namely (111) and (311) [belonging to $L2_1$] are not seen in the room temperature pattern. However, some additional reflections namely (201) and (12-5) [belonging to the martensite phase] can be seen with austenite (200) peak, which suggest that a fraction of martensite phase is still present at room temperature. In contrast, in the annealed ribbon, (111) and (311) peaks appear with noticeable intensity, as shown in figure 1(b). The interesting point to be noted is that these additional reflections could be seen after annealing the sample for as short a time as 1 hr. This result indicates that after annealing, the crystal structure changes to $L2_1$ from B2 phase. This implies that the short diffusion length resulting from the rapid solidification during the melt spinning process requires very little time to acquire the $L2_1$ ordered phase. In the case of bulk alloy of the same composition, the XRD pattern was identical to that of the annealed ribbon, at room temperature. In order to confirm the nature of ordering as derived from the XRD pattern, selected area electron diffraction patterns (SAEDP) were taken. The diffraction pattern as shown in figure 1(c) is taken from those few observed martensite phase patterns in the as-spun ribbon, which gives x/5{220} $L2_1$ reflection spots in between the main spots. As given by arrow marks it can be seen that the reciprocal lattice is divided into 5 equal parts by 4 satellite spots. These satellite spots could be illustrated as modulated 5 layered (5M) martensite or sometimes it is also termed as 10M martensite structure. In figure 1(d) austenite phase reflection spots can be observed, which correspond to austenite peaks. The diffraction pattern is taken in [011] zone axis at room temperature for the annealed ribbon. It is to be mentioned here that well defined diffraction spots corresponding to the austenite peaks have been found in [111] zone axis (not shown in the figure). However, in order to establish that the diffraction spots correspond to the odd peaks generated through XRD pattern, the SAEDP is taken in [011] zone axis and is shown in figure 1(d). The observed {111} diffraction spots can be ascribed to well defined $L2_1$ structure of the system. The $L2_1$ ordered phase from diffraction pattern is also reported in $Ni_2FeGa$ system [15]. The observation of well defined odd peaks in XRD pattern and the {111} spot in the diffraction pattern signify the increase in the degree of atomic order from B2 to $L2_1$ phase after annealing.

The composition analysis of bulk, as spun and annealed ribbon of NiCoMnSb alloy is presented in Table 1. It has been found that there are small changes in the composition among bulk, as spun ribbon and annealed ribbon, which are within the error bar.

Table 1: Composition analysis in bulk, as-spun ribbon and annealed ribbon of NiCoMnSb

| NiCoMnSb alloy | Atomic % with error | | | |
|---|---|---|---|---|
| | Ni | Co | Mn | Sb |
| Bulk | 44.60±0.94 | 3.88±0.51 | 39.46±0.62 | 12.06±0.28 |
| As Spun | 44.07±1.01 | 4.23±0.55 | 39.57±0.66 | 12.12±0.30 |
| Annealed | 44.45±1.02 | 4.26±0.56 | 39.95±0.67 | 11.34±0.31 |

The zero field cooled (ZFC) and field cooled cooling (FCC) magnetization curves measured in 1 kOe field for both the as-spun and the annealed samples are shown in figure 2(a). On decreasing the temperature there is a sudden decrease in the magnetization which corresponds to the first order phase transition from austenite to martensitic phase. The presence of hysteresis between the ZFC and FCC curves around the martensitic transition implies the first order nature of the transition, where the coexistence of both martensite and austenite phases is observed. The sudden fall in the magnetization indicates the presence of non-ferromagnetic components in the martensite phase. In the heating (ZFC) curve, the martensitic transition starts at $A_S$ (austenite start) and finishes at $A_F$ (austenite finish). Similarly in the cooling (FCC) curve martensitic transition starts at $M_S$ (martensite start) and finishes at $M_F$ (martensite finish). By comparing the M-T data of the as-spun and the annealed ribbons, it can be seen that the later one shows a sharper martensitic transition, indicating an enhanced magneto-structural coupling after annealing. It can be noticed that the magnetization value in both the austenite and the martensite phases increases after annealing. This suggests that the ferromagnetic order in the sample increases after annealing. It can also be observed that the martensitic transition shifts to lower temperature in the annealed ribbon as compared to the as-spun ribbon, which signifies the stabilization of austenite phase. As given in table 2 the martensitic transition temperature, $A_S$, for $L2_1$ phase is 15 K lower than the B2 phase. The corresponding transition temperature for the bulk alloy of the same composition is 292 K, which is larger than that of the as-spun ribbon by 7 K. The thermo-magnetic irreversibility between the ZFC and FCC curves at low temperature can be noticed in both the samples. The splitting between the ZFC and FCC curves indicates the magnetic frustration or re-entrant spin glass behavior [16], which arise due to the presence of both antiferromgnetic (AFM) and ferromagnetic interactions in the off-stoichiometric alloy. However, as shown in figure 2(a) and (b), the irreversibility starts at about 200 K in case of as as-spun ribbon, whereas it starts only close to 100 K in the case of the annealed sample. The disorder in Mn and Sb during quenching and annealing the ribbon, may affect on Mn-Mn exchange interaction. The competing magnetic interactions may lead to a decrease in the irreversibility temperature, which reinforces the assumption about the growth of the ferromagnetic phase with annealing. A comparison of various transition temperatures of the as-spun and the annealed ribbons along with those of the bulk alloy is shown in Table 2.

Table 2 Martensitic transition temperatures, exchange bias field, isothermal magnetic entropy change and magnetoresistance ratio in NiCoMnSb system.

| $Ni_{46}Co_4Mn_{38}Sb_{12}$ | $M_S$ (K) | $M_F$ (K) | $A_S$ (K) | $A_F$ (K) | $H_{EB}$ (Oe) | $\Delta S_M$ (J/kg K) | -MR(%) |
|---|---|---|---|---|---|---|---|
| As- spun | 296 | 270 | 285 | 310 | 416 | 17 | 10 |
| Annealed | 283 | 259 | 270 | 298 | 428 | 33 | 14 |
| Bulk | 295 | 287 | 292 | 304 | 450 | 40 | 28 |

To study the effect of rapid quenching on the hysteretic behavior around the martensitic transition, we have measured the M-H curves at different temperatures as shown in figure 2(b). The M-H loops for the annealed ribbons are shown in the inset of figure 2(b). It is found that the as-spun ribbon possesses less hysteresis than the annealed one. Moreover, the magnetization in the austenite phase is larger for the annealed ribbons than the as-spun ribbons. A magnetization difference (ΔM) of 22 emu/g is observed between FM austenite phase and nearly paramagnetic martensite phase in the case of the as-spun ribbon whereas for annealed sample it is found to be 28 emu/g. For bulk sample the ΔM value is observed as 30 emu/g [13].

The five loop M-H isotherms as shown in figure 3(a) are taken at 295 K and 286 K for the as-spun and annealed ribbons respectively. These two temperatures fall nearly in the highly metastable region as seen from the M-H curve. A field induced first order metamagnetic transition from lower magnetic, martensite phase to higher magnetic, austenite phase can be observed in both the as-spun and the annealed ribbons. However, the annealed ribbon shows a sharper metamagnetic transition than the as-spun one, which can be clearly seen from the dM/dH curves shown in the inset of figure 3(a). The slope change near the metamagnetic transition is related to magnetic and microstructural contributions. The rearrangement of martensitic twins, as a result of field induced de-twinning and the moment rotation are responsible for the slope change. The field induced twin boundary motion is also reported in NiMnGa alloy [17, 18]. However, the field required for the de-twinning in the present case is found to be larger than that of NiMnGa. Large field induced irreversibility in the M-H loops is due to kinetic arrest of the supercooled austenite phase in both as spun and annealed ribbon. In contrast, no metamagnetic transition can be found in the negative field regime for both the ribbons. The smaller hysteresis in the as-spun ribbon is related to weaker metamagnetic transition in that sample.

Figure 3(b) shows the field cooled magnetization behavior of the as-spun and the annealed ribbons at 3 K. The hysteresis loops are taken after cooling the sample in a field of 50 kOe. For the as-spun and annealed ribbons, the shift in loops, i.e., the exchange bias field ($H_{EB}$) is found to be 416 Oe and 428 Oe respectively at 3 K. This shift is attributed to the exchange bias effect, which occurs due to the presence of FM/ AFM mixed phase at low temperatures [19]. The existence of double shifted loops can be seen in the ZFC M-H plots in both the ribbons, as shown in the inset of figure 3(b), again confirming the coexistence of AFM and FM components in the system. The EB field reported in the bulk $Ni_{46}Co_4Mn_{38}Sb_{12}$ alloy is 450 Oe [13], which is nearly the same as that obtained in the annealed ribbon. The coercivity observed for the as-spun ribbon is 403 Oe at 3 K, which significantly reduces to 177 Oe upon annealing. The corresponding value for the bulk alloy of the same composition is 336 Oe [13]. The saturation magnetization ($M_S$) of the as-spun ribbon is 28 emu/g whereas that of annealed ribbon is 33 emu/g at 3 K. Therefore, increase in the magnetization and the reduction in the coercivity with annealing clearly reflects an enhancement of the ferromagnetic component in the sample.

Figure 4 shows the variation of isothermal magnetic entropy change ($\Delta S_M$) with temperature near the martensitic transition region, calculated using the Maxwell's relation [20]. For the as-spun $Ni_{46}Co_4Mn_{38}Sb_{12}$ ribbon, the maximum $\Delta S_M$ value is 16.7 J/kg K for a field change of 50 kOe, which significantly increases to 32.5 J/kg K for the annealed sample. The large ΔM value observed in the annealed sample has enhanced the $\Delta S_M$ value significantly. It may be noted that the entropy change for the bulk alloy is 32 J/kg K [12]. It is known that the Maxwell's relation is only valid near a second order transition for calculating magnetic entropy change. But it can be used near first order transition if the discontinuity at the transition is not very sharp. In order to confirm this, the $\Delta S_M$ value is also calculated in the decreasing field mode. In this mode the maximum entropy changes are observed as 17 J/kg K and 32.8 J/kg K for the as-spun and the annealed ribbons respectively. Therefore, the $\Delta S_M$ value observed for decreasing field mode is nearly same as that for the increasing field mode. Comparing the MCE results on the as-spun and the annealed ribbons, it can be pointed out that the magneto-structural coupling has improved after annealing. However increase in the magnetic hysteresis loss from 13 J/kg to 19 J/kg (for 50 kOe) after annealing is detrimental for the refrigerant capacity (RC). A rough estimate of the hysteresis loss has been obtained by calculating the area between the magnetization curves during increasing and decreasing field cycles. RC has been calculated by integrating $\Delta S_M$ (T) over the full width at half maximum of the $\Delta S_M$ (T)-T peak. The effective refrigeration capacity, which is obtained by subtracting the hysteresis loss from the calculated RC value, is found to be 60 and 55 J/kg at 50 kOe in the annealed and the as-spun ribbons respectively. It is to be noted that the RC value for the bulk material of the same composition is reported to be 95 J/kg. However, the larger hysteresis loss value of 21 J/kg was observed, which gives rise to an effective RC of 74 J/kg in the bulk alloy [12]. As the hysteresis loss plays a major role in designing a magnetic refrigerant, melt spun ribbons may offer a practical possibility. However, it has to be realized that the fields required for significant MCE is larger compared to that of many known potential refrigerant materials.

Since significant increase in magneto-structural coupling after annealing gives rise to enhanced MCE value, similar results are expected to reflect in the magnetoresistance as well. In the figure 5(a) and (b), 4-loop MR isotherms are shown in increasing and decreasing fields near martensitic transition region at 286, 295 and 300 K. It can be seen that a maximum MR value of -10% is obtained at 295 K for the as-spun ribbon. In the case of the annealed sample, a maximum value is about -14% near 286 K. It is of importance to note that the MR isotherms show large hysteresis and also that the zero field values at the beginning and at the end of the field cycle are quite different. This field-induced irreversibility is a characteristic of the metastability of the two competing structural (and magnetic) phases. In the case of as-spun ribbon, an increase in temperature above 295 K causes the metastability to decrease, which is accompanied by a decrease of MR to -7% at 300 K. In contrast, as the martensitic transition for the annealed ribbon is sharper, at 295 K and 300 K the sample is already in the austenite phase. Therefore, at these temperatures the MR shows normal metallic behavior with no hysteresis (Fig. 5b).

Based on the above results, it is clear that melt spinning as well as annealing affect the structural, magnetic, magnetocaloric and magneto-transport properties of this alloy. $L2_1$ cubic structure is observed at room temperature for the bulk alloy. But for melt spun ribbon the structure is found to be B2 type which transforms to $L2_1$ ordered phase upon annealing. As a result of this stabilization of austenite phase, the martensitic transition shifts towards lower temperatures after annealing. One of the very interesting outcomes of the present work is that most of the results obtained in the annealed ribbon approach those of the bulk $Ni_{46}Co_4Mn_{38}Sb_{12}$ system, even though the time of annealing was only 1 hour. In the bulk system MCE value of 32 J/kg K and MR value of -28% have been obtained near the martensitic transition [12, 14]. After annealing for 1 hr there is a significant increase in MCE, MR and EB values. The obtained MCE value for the annealed ribbon is close to the value obtained from the bulk sample [12]. Present study reveals that in the bulk form, the magnetic hysteresis is larger than that of the annealed ribbon near the martensitic transition, which causes large hysteresis loss in the bulk. Thus, as compared to the bulk, ribbons may be more suitable as magnetic refrigerants. It is to be noted that the bulk sample of same composition also exhibits $L2_1$ ordering in the austenite phase. However, it is also of importance that there are some differences between the bulk alloy and the annealed ribbon. The martensitic transition temperature $[T_M=M_S+M_F)/2]$ of the bulk alloy is about 288 K, whereas it is 283 K for the as-spun ribbon and 271 K for the annealed ribbon. During rapid quenching process, stress is introduced in melt spun ribbon, as a result of which lower martensitic transition temperature is observed in the ribbon as compared to the bulk [21]. This can be attributed to the difference in the $L2_1$ phase in the bulk and the annealed ribbon. It has been noted that the odd reflections as seen from XRD and FEG-TEM are much weaker in annealed ribbon, as compared to those in the bulk. This signifies the presence of some amount of Mn-Sb disorder in the $L2_1$ phase. Some more detailed microscopic experiments are needed to confirm this disorder, which is expected to modify the exchange interactions in the ribbon.

## 4. Conclusions

In conclusion, we have studied the structural, magnetic, magnetocaloric and magnetoresistance properties of as-spun ribbon of $Ni_{46}Co_4Mn_{38}Sb_{12}$ and have compared them with the results obtained on the annealed ribbon as well as the bulk alloy of the same composition. From XRD it is found that the degree of atomic order or $L2_1$ order increases after annealing. This results in the stabilization of the austenite phase in the annealed ribbon, thereby reducing the martensitic transition temperature. The FM order increases after annealing, as revealed by the increase in the magnetization of austenite as well as the martensite phases. The coercivity value at 3 K is also found to decrease significantly after annealing. Increase in the magneto-structural coupling after annealing causes an enhancement of MCE and MR values considerably. Most of the properties of the annealed ribbon are nearly similar to those of the bulk. The minor differences between them are attributed to some disorder within the $L2_1$ phase. It is possible that a longer annealing time may fully restore the properties to those of the bulk. But it is quite clear that the multifunctional properties of this alloy have undergone an overall improvement as a result of melt spinning and annealing. In summary, the present study shows that melt spinning and annealing technique offer another tool to tune the magnetic and other related properties of Heusler alloys.

**Acknowledgement:**

The authors thank SAIF and MEMS, IIT Bombay for their help in collecting the microstructural data. They are also grateful to Devendra Buddhikot for his help in the resistivity measurements. KGS acknowledges the financial support received from CSIR, Govt of India.

**References:**

[1] K. Ullakko, J. K. Huang, C. Kantner, R. C. O. Handley, V. V. Kokorin, Applied Physics Letter 69 (1996) 1966.

[2] T. Krenke, E. Duman, M. Acet, E. F. Wassermann, X. Moya, L. Mañosa, A. Planes, Nature 4 (2005) 450.

[3] V. K. Pecharsky, K. A. Gschneidner, Journal of Magnetism and Magnetic Materials 167 (1997) L179.

[4] S. Y. Yu, L. Ma, G. D. Liu, J. L. Chen, Z. X. Cao, G. H. Wu, B. Zhang, X. X. Zhang, Applied Physics Letter 90 (2007) 242501.

[5] V. K. Sharma, M. K. Chattopadhyay, K. H. B. Shaeb, A. Chouhan, S. B. Roy, Applied Physics Letter 89 (2006) 2225091.

[6] J. Liu, N. Scheerbaum, D. Hinz, O. Gutfliesch, Acta Materialia 56 (2008) 3177.

[7] Y. Ma, C. Jiang, H. Xu, C. Wang, X. Liu, Acta Materialia 55 (2007) 1533.

[8] D. M. Raj Kumar, N. V. Rama Rao, M. Manivel Raja, D. V. Sridhara Rao, M. Srinivas, S. Esakki Muthu, S. Arumugam, K. G. Suresh, Journal of Magnetism and Magnetic Materials 324 (2012) 26.

[9] C. Liu, W. Zhang, Z. Qian, Z. Hua, Q. Zhao, Y. Sui, W. Su, M. Zhang, Z. Liu, G. Liu, G. Wu, Journal of Alloys and Compounds 433 (2007) 37.

[10] V. Sanchez- Alarcos, J. I. Prez-Landazabal, V. Recarte, J. A. Rodriguez-Velamazan, V. A. Cherneko, Journal of Physics: Condensed Matter 22 (2010) 166001.

[11] R. Santamarta, E. Cesari, J. Font, J. Muntasell, J. Pons, J. Dutkiewicz, Scripta Materialia 54 (2006) 1985.

[12] A. K. Nayak, K. G. Suresh, A. K. Nigam, A. A. Coelho, S. Gama, Journal of Applied Physics 106 (2009) 053901.

[13] A. K. Nayak, K. G. Suresh, A. K. Nigam, Journal of Physics D: Applied Physics 42 (2009) 115004.

[14] A. K. Nayak, K. G. Suresh, A. K. Nigam, Journal of Applied Physics 107 (2010) 09A9271.

[15] H. R. Zhang, C. Ma, H. F. Tian, G. H. Wu, J. Q. Li, Physical Review B 77 (2008) 214106.

[16] A. K. Nayak, K. G. Suresh, A. K. Nigam, Journal of Physics: Condensed Matter 23 (2011) 416004.

[17] J. Gutiérrez, J. M. Barandiar´, P. Lázpita, C. Segui, E. Cesari, Sensors and Actuators A 129 (2006) 163.

[18] P. Lázpita, J. Gutiérrez, J. M. Barandiar´, R. C. O'Handley, Sensors Letters 5(1) (2007) 65.

[19] M. Khan, I. Dubenku, S. Stadler, N. Ali, Applied Physics Letter 91 (2007) 072510.

[20] K. A. Gschneidner, V. K. Pecharsky, A. O. Tsokol, Reports on Progress in Physics 68 (2005) 1479.

[21] E. Cesari, P. Ochin, R. Portier, V. Kolomytsev, Yu. Koval, A. Pasko, V. Soolshenko, Material Science & Engineering A 273-275 (1999) 738-744.

**Figures Captions:**

Fig. 1. XRD patterns of (a) as-spun and (b) annealed $Ni_{46}Co_4Mn_{38}Sb_{12}$ alloy ribbons recorded at room temperature. (c) and (d) show the SAEDP obtained for the as- spun and the annealed ribbons.

Table.1. Compositions of $Ni_{46}Co_4Mn_{38}Sb_{12}$ alloy (in the form of bulk, as-spun ribbon and annealed ribbon) determined by EDX analysis in atomic %.

Fig. 2. (a) ZFC and FCC magnetization curves as a function of temperature in 1 kOe for the as the as-spun and the annealed ribbons. (b) Isothermal magnetization curves in the vicinity of martensitic transition region for the as-spun (main panel) and the annealed ribbons (in the inset) of $Ni_{46}Co_4Mn_{38}Sb_{12}$.

Table.2. Various martensitic transition temperatures, exchange bias field at 3 K, maximum magnetic entropy change and maximum magnetoresistance values in the as-spun and the annealed ribbons along with the bulk $Ni_{46}Co_4Mn_{38}Sb_{12}$ alloy.

Fig. 3(a). 5 loop field dependence of magnetization for the as-spun ribbon at 295 K and for the annealed ribbon at 286 K (b) Field cooled hysteresis loops of the as-spun and the annealed ribbons of $Ni_{46}Co_4Mn_{38}Sb_{12}$ alloy at 3 K. The inset in (a) shows the first derivative of the magnetization with field and that in (b) shows the ZFC M-H plots at 3 K.

Fig. 4. Temperature variation of isothermal magnetic entropy change ($\Delta S_M$) in increasing (solid line) and decreasing field (dotted line) modes in the as-spun and the annealed $Ni_{46}Co_4Mn_{38}Sb_{12}$ ribbons.

Fig. 5. Field variation of magnetoresistance for (a) as-spun ribbon and (b) annealed ribbon of $Ni_{46}Co_4Mn_{38}Sb_{12}$

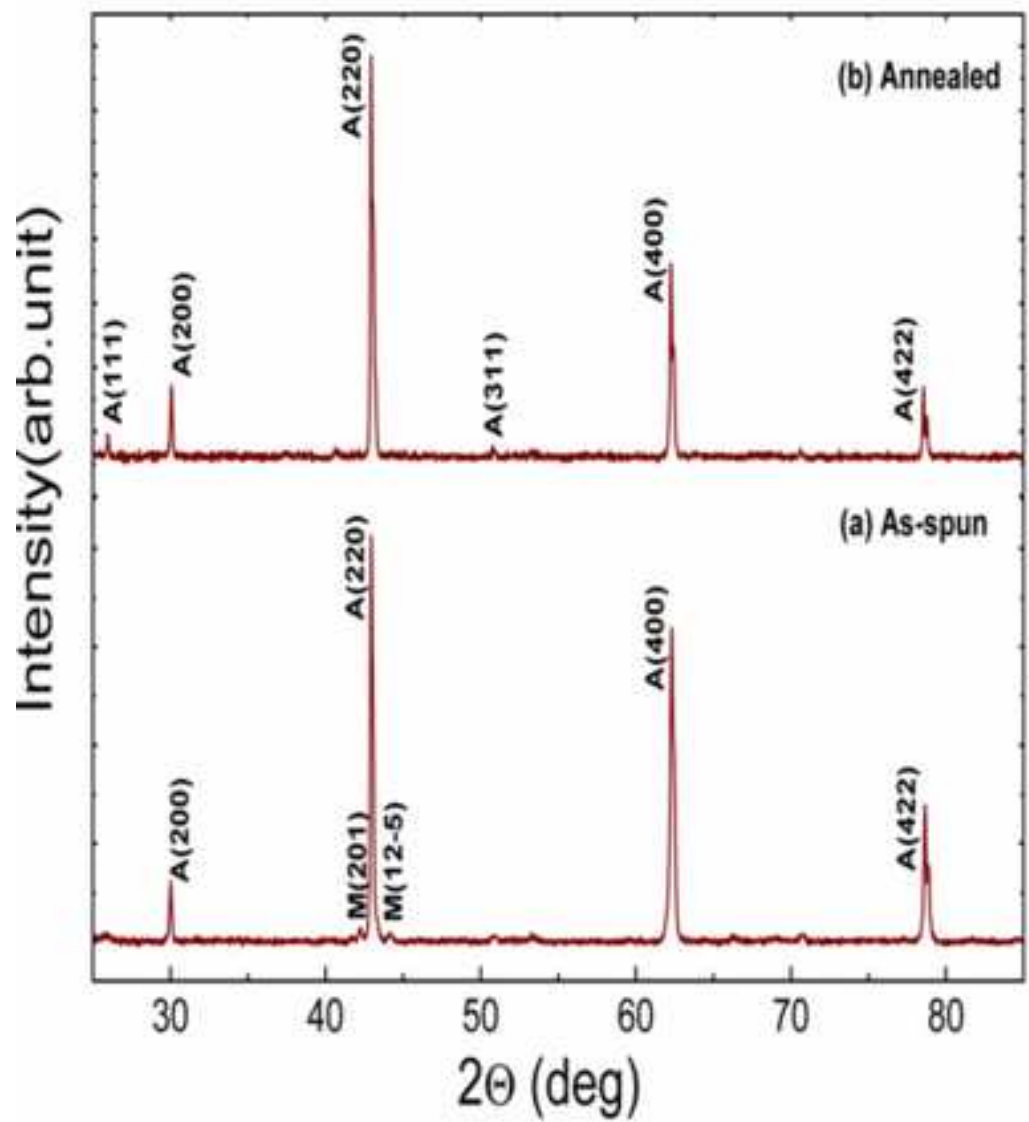
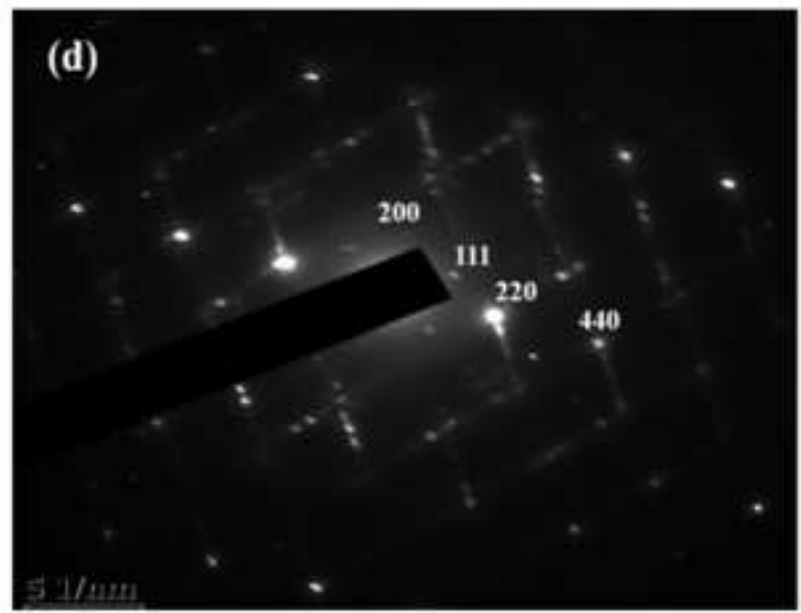
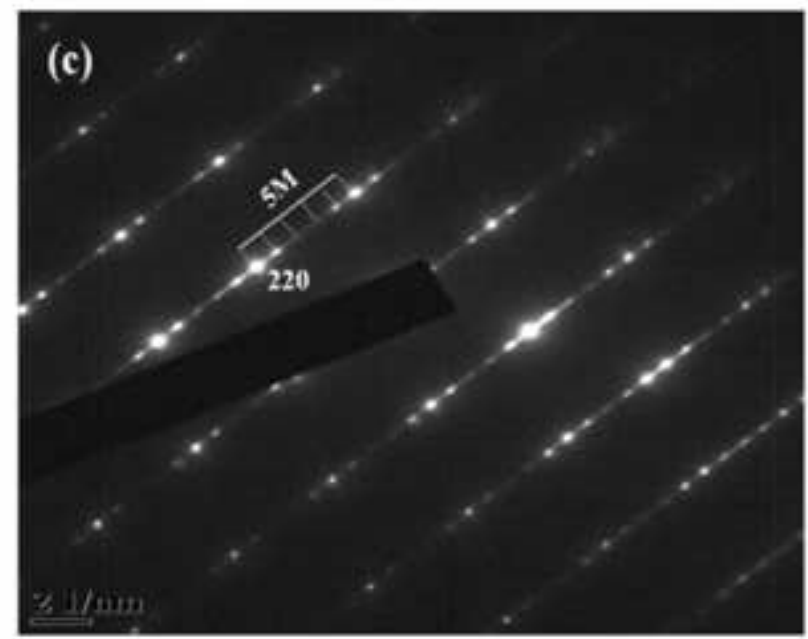

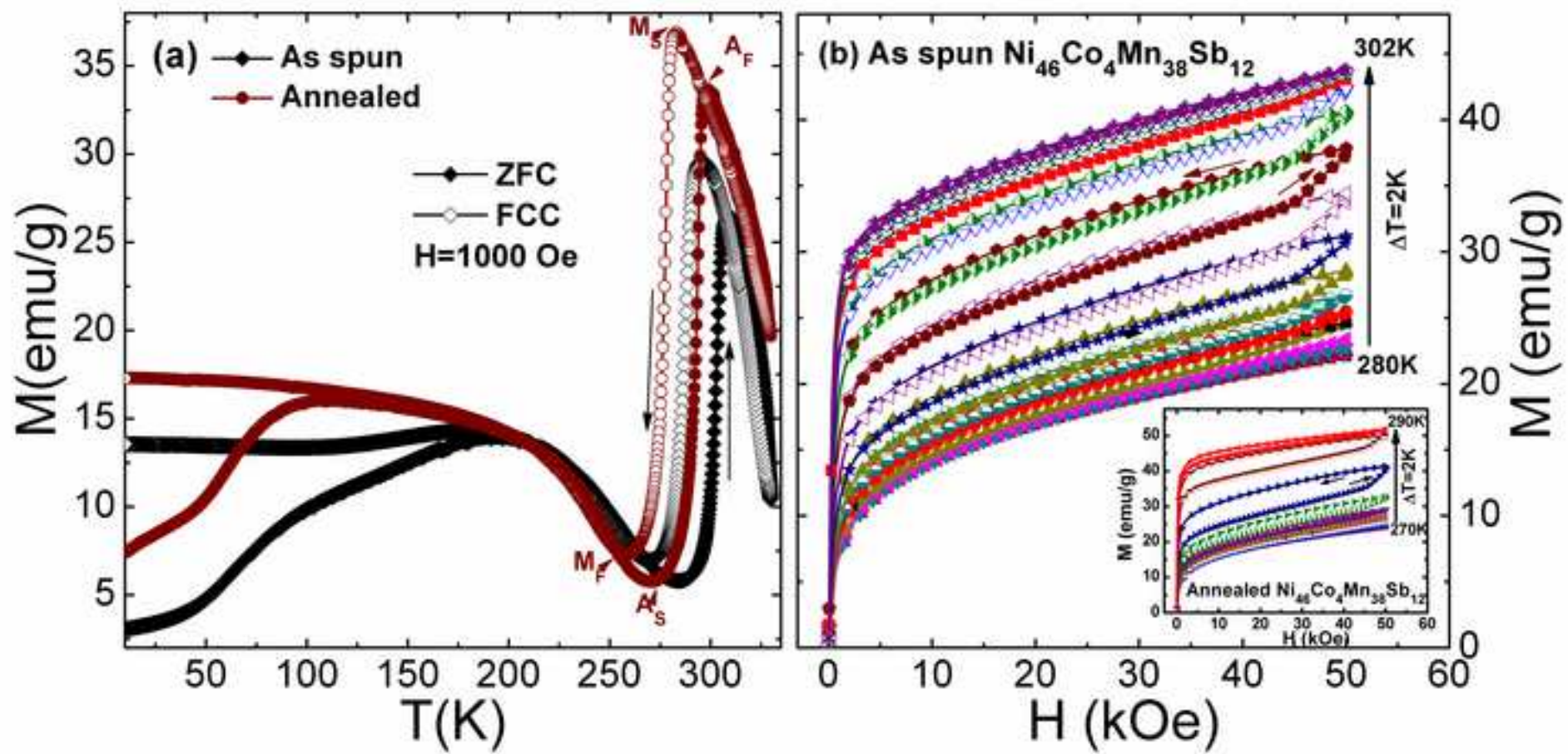

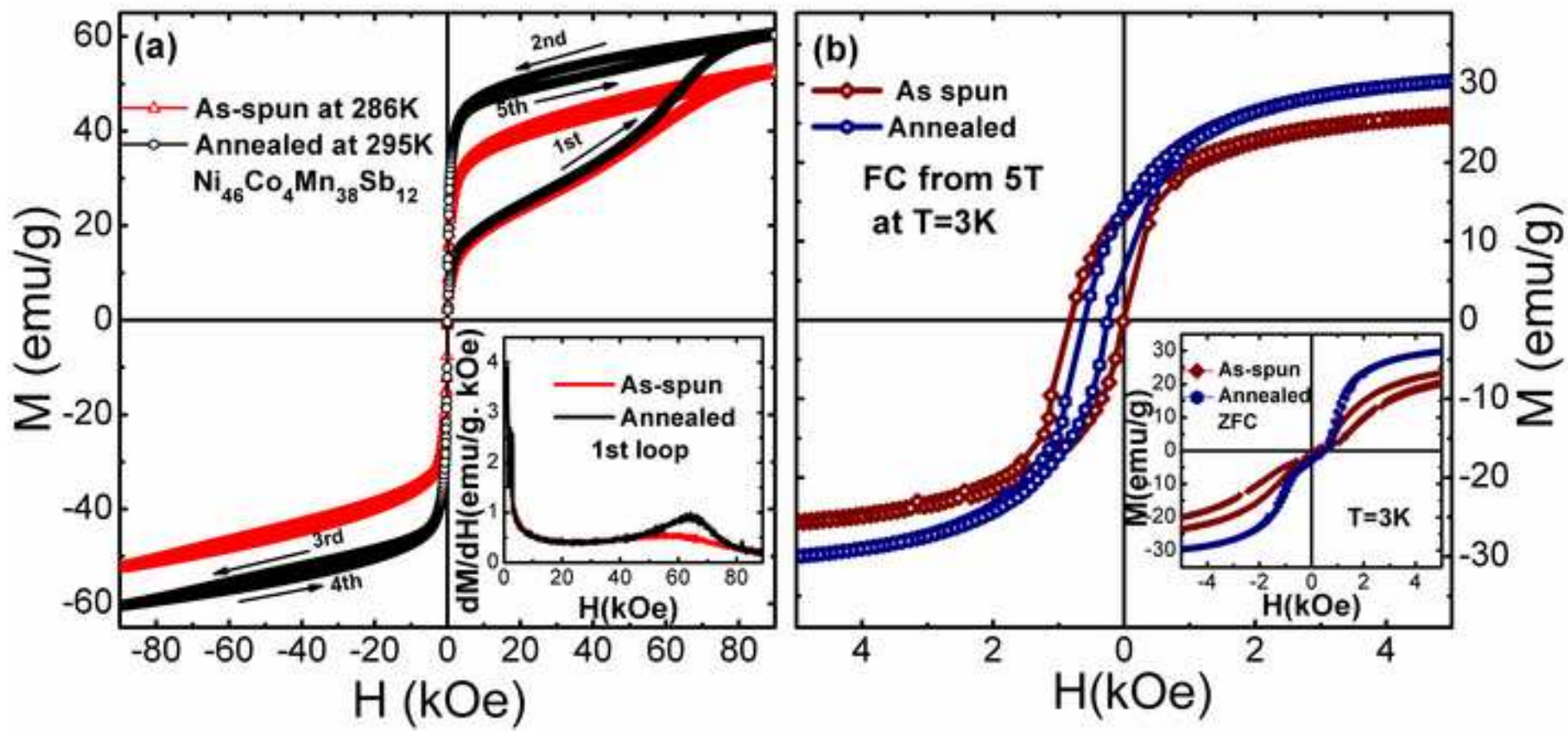

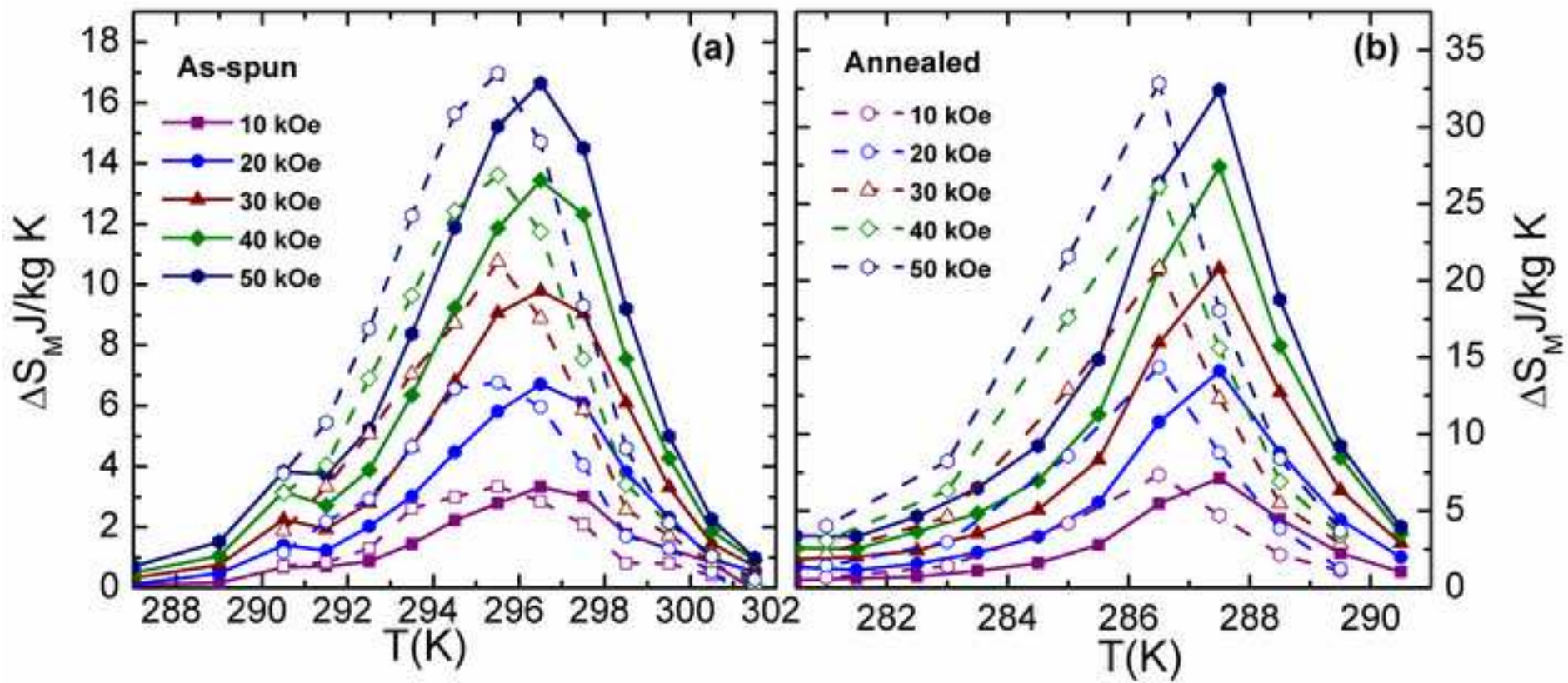

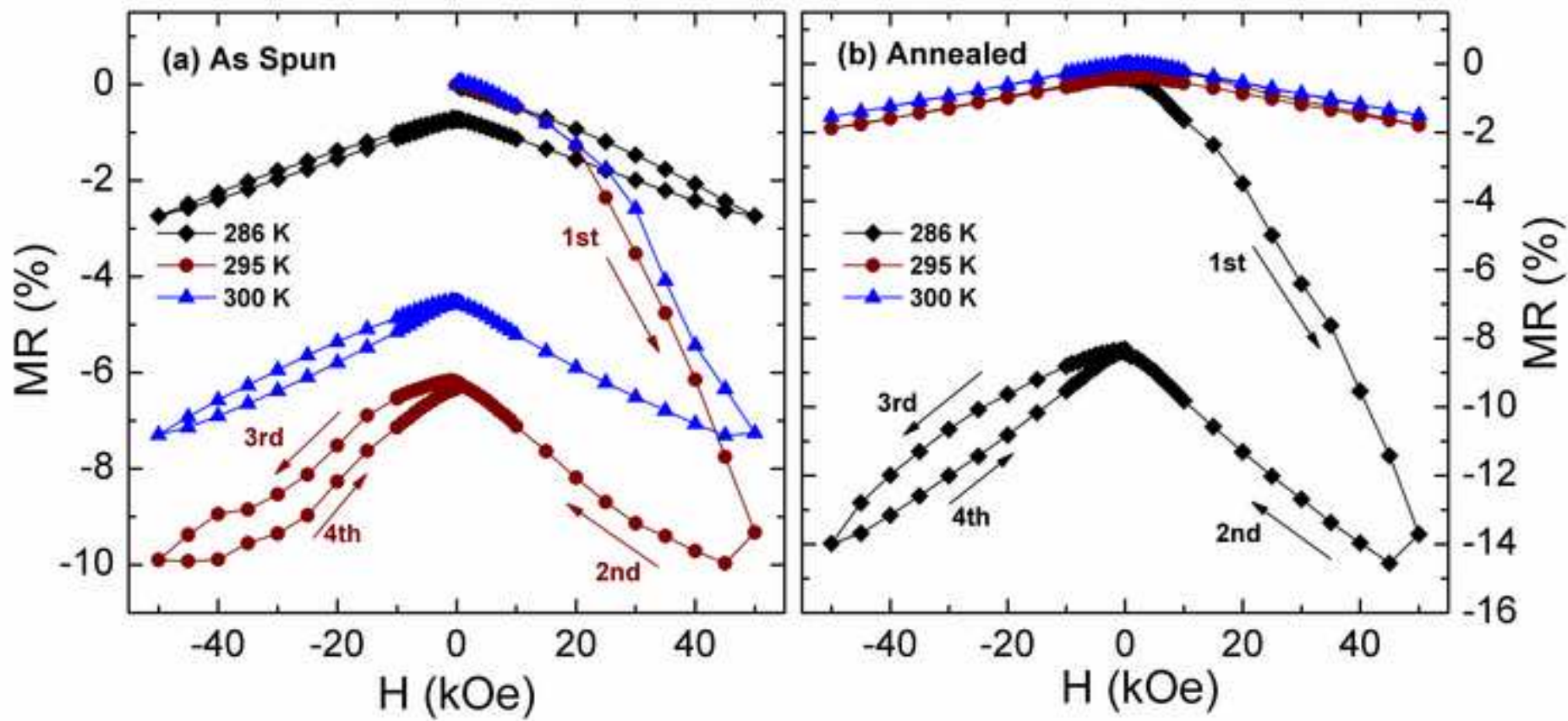

**Table**

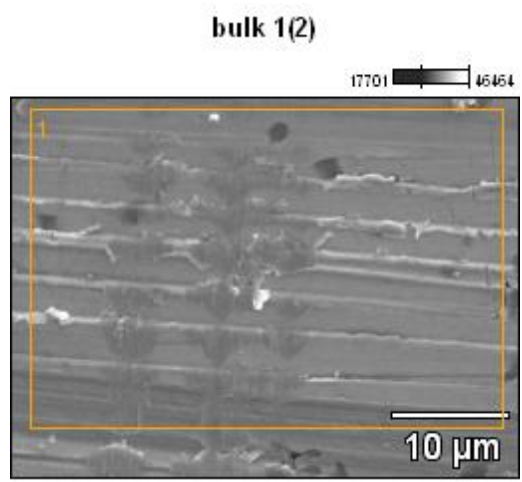

Image Name: bulk 1(2)

Accelerating Voltage: 15.0 kV

Magnification: 3000

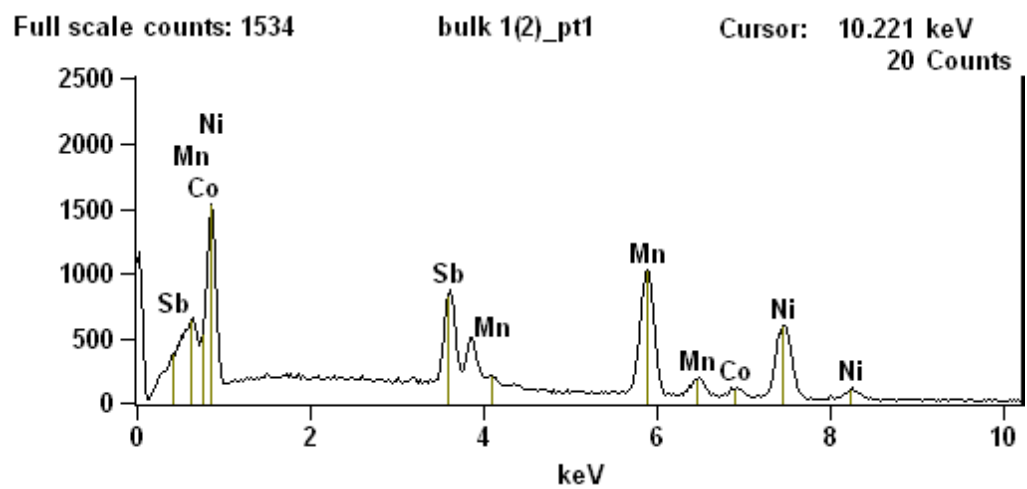

## Net Counts

|               | Mn     | Co    | Ni     | Sb     |
|---------------|--------|-------|--------|--------|
| bulk 1(2)_pt1 | 18647  | 1327  | 12500  | 18166  |

## Net Counts Error

|               | Mn     | Co    | Ni     | Sb     |
|---------------|--------|-------|--------|--------|
| bulk 1(2)_pt1 | +/-291 | +/-175| +/-264 | +/-428 |

## Weight %

|               | Mn    | Co   | Ni    | Sb    |
|---------------|-------|------|-------|-------|
| bulk 1(2)_pt1 | 33.43 | 3.53 | 40.38 | 22.65 |

## Weight % Error

|               | Mn      | Co      | Ni      | Sb      |
|---------------|---------|---------|---------|---------|
| bulk 1(2)_pt1 | +/-0.52 | +/-0.47 | +/-0.85 | +/-0.53 |

## Atom %

|               | Mn    | Co   | Ni    | Sb    |
|---------------|-------|------|-------|-------|
| bulk 1(2)_pt1 | 39.46 | 3.88 | 44.60 | 12.06 |

## Atom % Error

|               | Mn      | Co      | Ni      | Sb      |
|---------------|---------|---------|---------|---------|
| bulk 1(2)_pt1 | +/-0.62 | +/-0.51 | +/-0.94 | +/-0.28 |

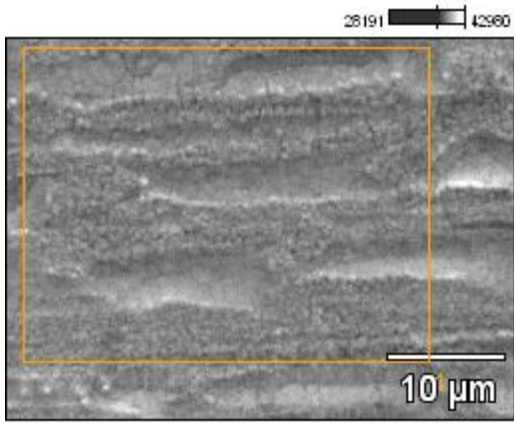

Image Name: ribbon as spun(2)

Accelerating Voltage: 15.0 kV

Magnification: 3000

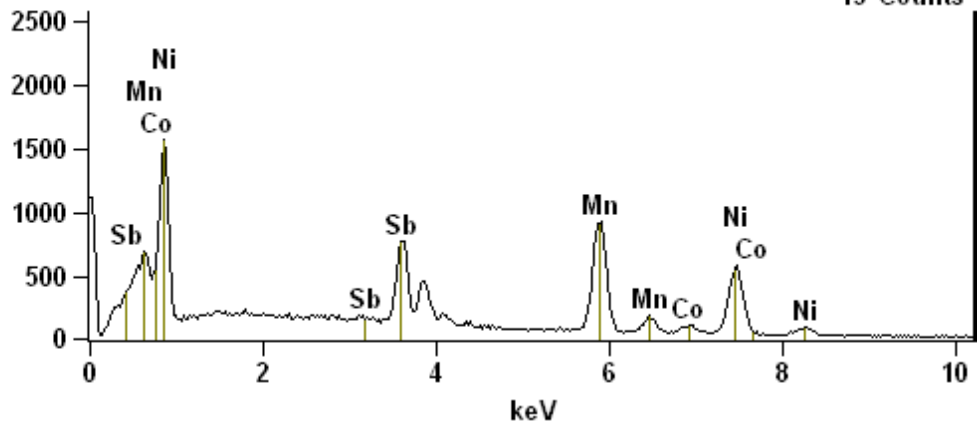

| Net Counts | | | | |
|---|---|---|---|---|
| | *Mn* | *Co* | *Ni* | *Sb* |
| *ribbon as spun(2)_pt1* | 16673 | 1289 | 11016 | 16275 |

| Net Counts Error | | | | |
|---|---|---|---|---|
| | *Mn* | *Co* | *Ni* | *Sb* |
| *ribbon as spun(2)_pt1* | +/-276 | +/-169 | +/-252 | +/-407 |

| Weight % | | | | |
|---|---|---|---|---|
| | *Mn* | *Co* | *Ni* | *Sb* |
| *ribbon as spun(2)_pt1* | 33.52 | 3.84 | 39.89 | 22.75 |

| Weight % Error | | | | |
|---|---|---|---|---|
| | *Mn* | *Co* | *Ni* | *Sb* |
| *ribbon as spun(2)_pt1* | +/-0.55 | +/-0.50 | +/-0.91 | +/-0.57 |

| Atom % | | | | |
|---|---|---|---|---|
| | *Mn* | *Co* | *Ni* | *Sb* |
| *ribbon as spun(2)_pt1* | 39.57 | 4.23 | 44.07 | 12.12 |

| Atom % Error | | | | |
|---|---|---|---|---|
| | *Mn* | *Co* | *Ni* | *Sb* |
| *ribbon as spun(2)_pt1* | +/-0.66 | +/-0.55 | +/-1.01 | +/-0.30 |

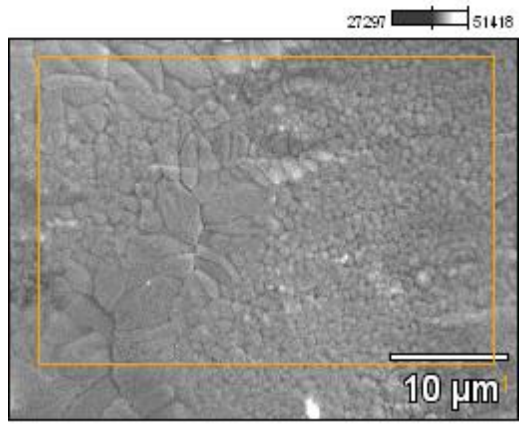

ribbon 1st annld(1)

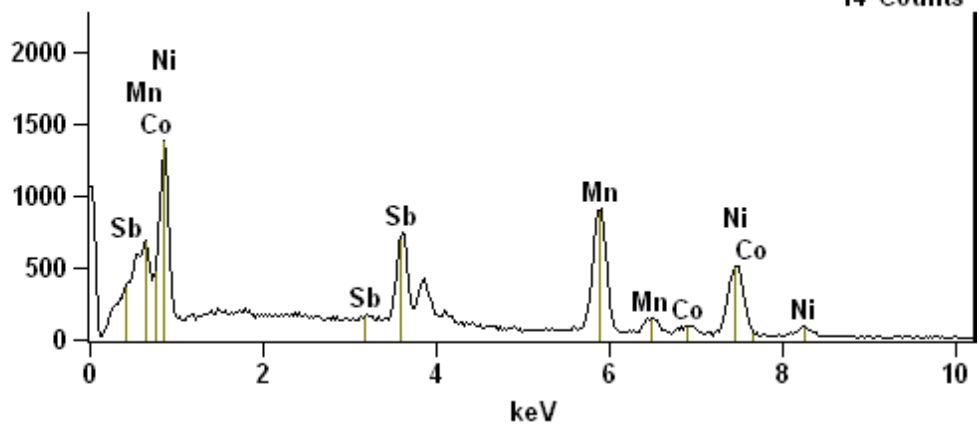

Full scale counts: 1383    ribbon 1st annld(1)_pt1    Cursor: 10.221 keV
14 Counts

| Net Counts | | | | |
|---|---|---|---|---|
| | *Mn* | *Co* | *Ni* | *Sb* |
| *ribbon 1st annld(1)_pt1* | 16152 | 1242 | 10633 | 14575 |

| Net Counts Error | | | | |
|---|---|---|---|---|
| | *Mn* | *Co* | *Ni* | *Sb* |
| *ribbon 1st annld(1)_pt1* | +/-272 | +/-164 | +/-244 | +/-394 |

| Weight % | | | | |
|---|---|---|---|---|
| | *Mn* | *Co* | *Ni* | *Sb* |
| *ribbon 1st annld(1)_pt1* | 34.10 | 3.90 | 40.54 | 21.45 |

| Weight % Error | | | | |
|---|---|---|---|---|
| | *Mn* | *Co* | *Ni* | *Sb* |
| *ribbon 1st annld(1)_pt1* | +/-0.57 | +/-0.52 | +/-0.93 | +/-0.58 |

| Atom % | | | | |
|---|---|---|---|---|
| | *Mn* | *Co* | *Ni* | *Sb* |
| *ribbon 1st annld(1)_pt1* | 39.95 | 4.26 | 44.45 | 11.34 |

| Atom % Error | | | | |
|---|---|---|---|---|
| | *Mn* | *Co* | *Ni* | *Sb* |
| *ribbon 1st annld(1)_pt1* | +/-0.67 | +/-0.56 | +/-1.02 | +/-0.31 |